
%


\catcode `\@=11 

\def\@version{1.3}
\def\@verdate{28.11.1992}


%
%
%
%
%
%

\font\fiverm=cmr5
\font\fivei=cmmi5	\skewchar\fivei='177
\font\fivesy=cmsy5	\skewchar\fivesy='60
\font\fivebf=cmbx5

\font\sevenrm=cmr7
\font\seveni=cmmi7	\skewchar\seveni='177
\font\sevensy=cmsy7	\skewchar\sevensy='60
\font\sevenbf=cmbx7

\font\eightrm=cmr8
\font\eightbf=cmbx8
\font\eightit=cmti8
\font\eighti=cmmi8			\skewchar\eighti='177
\font\eightmib=cmmib10 at 8pt	\skewchar\eightmib='177
\font\eightsy=cmsy8			\skewchar\eightsy='60
\font\eightsyb=cmbsy10 at 8pt	\skewchar\eightsyb='60
\font\eightsl=cmsl8
\font\eighttt=cmtt8			\hyphenchar\eighttt=-1
\font\eightcsc=cmcsc10 at 8pt
\font\eightsf=cmss8

\font\ninerm=cmr9
\font\ninebf=cmbx9
\font\nineit=cmti9
\font\ninei=cmmi9			\skewchar\ninei='177
\font\ninemib=cmmib10 at 9pt	\skewchar\ninemib='177
\font\ninesy=cmsy9			\skewchar\ninesy='60
\font\ninesyb=cmbsy10 at 9pt	\skewchar\ninesyb='60
\font\ninesl=cmsl9
\font\ninett=cmtt9			\hyphenchar\ninett=-1
\font\ninecsc=cmcsc10 at 9pt
\font\ninesf=cmss9

\font\tenrm=cmr10
\font\tenbf=cmbx10
\font\tenit=cmti10
\font\teni=cmmi10		\skewchar\teni='177
\font\tenmib=cmmib10	\skewchar\tenmib='177
\font\tensy=cmsy10		\skewchar\tensy='60
\font\tensyb=cmbsy10	\skewchar\tensyb='60
\font\tenex=cmex10
\font\tensl=cmsl10
\font\tentt=cmtt10		\hyphenchar\tentt=-1
\font\tencsc=cmcsc10
\font\tensf=cmss10

\font\elevenrm=cmr10 scaled \magstephalf
\font\elevenbf=cmbx10 scaled \magstephalf
\font\elevenit=cmti10 scaled \magstephalf
\font\eleveni=cmmi10 scaled \magstephalf	\skewchar\eleveni='177
\font\elevenmib=cmmib10 scaled \magstephalf	\skewchar\elevenmib='177
\font\elevensy=cmsy10 scaled \magstephalf	\skewchar\elevensy='60
\font\elevensyb=cmbsy10 scaled \magstephalf	\skewchar\elevensyb='60
\font\elevensl=cmsl10 scaled \magstephalf
\font\eleventt=cmtt10 scaled \magstephalf	\hyphenchar\eleventt=-1
\font\elevencsc=cmcsc10 scaled \magstephalf
\font\elevensf=cmss10 scaled \magstephalf

\font\fourteenrm=cmr10 scaled \magstep2
\font\fourteenbf=cmbx10 scaled \magstep2
\font\fourteenit=cmti10 scaled \magstep2
\font\fourteeni=cmmi10 scaled \magstep2		\skewchar\fourteeni='177
\font\fourteenmib=cmmib10 scaled \magstep2	\skewchar\fourteenmib='177
\font\fourteensy=cmsy10 scaled \magstep2	\skewchar\fourteensy='60
\font\fourteensyb=cmbsy10 scaled \magstep2	\skewchar\fourteensyb='60
\font\fourteensl=cmsl10 scaled \magstep2
\font\fourteentt=cmtt10 scaled \magstep2	\hyphenchar\fourteentt=-1
\font\fourteencsc=cmcsc10 scaled \magstep2
\font\fourteensf=cmss10 scaled \magstep2

\font\seventeenrm=cmr10 scaled \magstep3
\font\seventeenbf=cmbx10 scaled \magstep3
\font\seventeenit=cmti10 scaled \magstep3
\font\seventeeni=cmmi10 scaled \magstep3	\skewchar\seventeeni='177
\font\seventeenmib=cmmib10 scaled \magstep3	\skewchar\seventeenmib='177
\font\seventeensy=cmsy10 scaled \magstep3	\skewchar\seventeensy='60
\font\seventeensyb=cmbsy10 scaled \magstep3	\skewchar\seventeensyb='60
\font\seventeensl=cmsl10 scaled \magstep3
\font\seventeentt=cmtt10 scaled \magstep3	\hyphenchar\seventeentt=-1
\font\seventeencsc=cmcsc10 scaled \magstep3
\font\seventeensf=cmss10 scaled \magstep3

\def\@typeface{Computer Modern} 

\def\hexnumber@#1{\ifnum#1<10 \number#1\else
 \ifnum#1=10 A\else\ifnum#1=11 B\else\ifnum#1=12 C\else
 \ifnum#1=13 D\else\ifnum#1=14 E\else\ifnum#1=15 F\fi\fi\fi\fi\fi\fi\fi}

\def\mib{\hexnumber@\mibfam}
\def\syb{\hexnumber@\sybfam}

\def\makestrut{%
  \setbox\strutbox=\hbox{%
    \vrule height.7\baselineskip depth.3\baselineskip width 0pt}%
}

\def\bls#1{%
  \normalbaselineskip=#1%
  \normalbaselines%
  \makestrut%
}

%

\newfam\mibfam 
\newfam\sybfam 
\newfam\scfam  
\newfam\sffam  

\def\em{\ifdim\fontdimen1\font>0 \rm\else\it\fi}

\textfont3=\tenex
\scriptfont3=\tenex
\scriptscriptfont3=\tenex

\def\eightpoint{
  \def\rm{\fam0\eightrm}%
  \textfont0=\eightrm \scriptfont0=\sevenrm \scriptscriptfont0=\fiverm%
  \textfont1=\eighti  \scriptfont1=\seveni  \scriptscriptfont1=\fivei%
  \textfont2=\eightsy \scriptfont2=\sevensy \scriptscriptfont2=\fivesy%
  \textfont\itfam=\eightit\def\it{\fam\itfam\eightit}%
  \textfont\bffam=\eightbf%
    \scriptfont\bffam=\sevenbf%
      \scriptscriptfont\bffam=\fivebf%
  \def\bf{\fam\bffam\eightbf}%
  \textfont\slfam=\eightsl\def\sl{\fam\slfam\eightsl}%
  \textfont\ttfam=\eighttt\def\tt{\fam\ttfam\eighttt}%
  \textfont\scfam=\eightcsc\def\sc{\fam\scfam\eightcsc}%
  \textfont\sffam=\eightsf\def\sf{\fam\sffam\eightsf}%
  \textfont\mibfam=\eightmib%
  \textfont\sybfam=\eightsyb%
  \bls{10pt}%
}

\def\ninepoint{
  \def\rm{\fam0\ninerm}%
  \textfont0=\ninerm \scriptfont0=\sevenrm \scriptscriptfont0=\fiverm%
  \textfont1=\ninei  \scriptfont1=\seveni  \scriptscriptfont1=\fivei%
  \textfont2=\ninesy \scriptfont2=\sevensy \scriptscriptfont2=\fivesy%
  \textfont\itfam=\nineit\def\it{\fam\itfam\nineit}%
  \textfont\bffam=\ninebf%
    \scriptfont\bffam=\sevenbf%
      \scriptscriptfont\bffam=\fivebf%
  \def\bf{\fam\bffam\ninebf}%
  \textfont\slfam=\ninesl\def\sl{\fam\slfam\ninesl}%
  \textfont\ttfam=\ninett\def\tt{\fam\ttfam\ninett}%
  \textfont\scfam=\ninecsc\def\sc{\fam\scfam\ninecsc}%
  \textfont\sffam=\ninesf\def\sf{\fam\sffam\ninesf}%
  \textfont\mibfam=\ninemib%
  \textfont\sybfam=\ninesyb%
  \bls{12pt}%
}

\def\tenpoint{
  \def\rm{\fam0\tenrm}%
  \textfont0=\tenrm \scriptfont0=\sevenrm \scriptscriptfont0=\fiverm%
  \textfont1=\teni  \scriptfont1=\seveni  \scriptscriptfont1=\fivei%
  \textfont2=\tensy \scriptfont2=\sevensy \scriptscriptfont2=\fivesy%
  \textfont\itfam=\tenit\def\it{\fam\itfam\tenit}%
  \textfont\bffam=\tenbf%
    \scriptfont\bffam=\sevenbf%
      \scriptscriptfont\bffam=\fivebf%
  \def\bf{\fam\bffam\tenbf}%
  \textfont\slfam=\tensl\def\sl{\fam\slfam\tensl}%
  \textfont\ttfam=\tentt\def\tt{\fam\ttfam\tentt}%
  \textfont\scfam=\tencsc\def\sc{\fam\scfam\tencsc}%
  \textfont\sffam=\tensf\def\sf{\fam\sffam\tensf}%
  \textfont\mibfam=\tenmib%
  \textfont\sybfam=\tensyb%
  \bls{12pt}%
}

\def\elevenpoint{
  \def\rm{\fam0\elevenrm}%
  \textfont0=\elevenrm \scriptfont0=\eightrm \scriptscriptfont0=\fiverm%
  \textfont1=\eleveni  \scriptfont1=\eighti  \scriptscriptfont1=\fivei%
  \textfont2=\elevensy \scriptfont2=\eightsy \scriptscriptfont2=\fivesy%
  \textfont\itfam=\elevenit\def\it{\fam\itfam\elevenit}%
  \textfont\bffam=\elevenbf%
    \scriptfont\bffam=\eightbf%
      \scriptscriptfont\bffam=\fivebf%
  \def\bf{\fam\bffam\elevenbf}%
  \textfont\slfam=\elevensl\def\sl{\fam\slfam\elevensl}%
  \textfont\ttfam=\eleventt\def\tt{\fam\ttfam\eleventt}%
  \textfont\scfam=\elevencsc\def\sc{\fam\scfam\elevencsc}%
  \textfont\sffam=\elevensf\def\sf{\fam\sffam\elevensf}%
  \textfont\mibfam=\elevenmib%
  \textfont\sybfam=\elevensyb%
  \bls{13pt}%
}

\def\fourteenpoint{
  \def\rm{\fam0\fourteenrm}%
  \textfont0\fourteenrm  \scriptfont0\tenrm  \scriptscriptfont0\sevenrm%
  \textfont1\fourteeni   \scriptfont1\teni   \scriptscriptfont1\seveni%
  \textfont2\fourteensy  \scriptfont2\tensy  \scriptscriptfont2\sevensy%
  \textfont\itfam=\fourteenit\def\it{\fam\itfam\fourteenit}%
  \textfont\bffam=\fourteenbf%
    \scriptfont\bffam=\tenbf%
      \scriptscriptfont\bffam=\sevenbf%
  \def\bf{\fam\bffam\fourteenbf}%
  \textfont\slfam=\fourteensl\def\sl{\fam\slfam\fourteensl}%
  \textfont\ttfam=\fourteentt\def\tt{\fam\ttfam\fourteentt}%
  \textfont\scfam=\fourteencsc\def\sc{\fam\scfam\fourteencsc}%
  \textfont\sffam=\fourteensf\def\sf{\fam\sffam\fourteensf}%
  \textfont\mibfam=\fourteenmib%
  \textfont\sybfam=\fourteensyb%
  \bls{17pt}%
}

\def\seventeenpoint{
  \def\rm{\fam0\seventeenrm}%
  \textfont0\seventeenrm  \scriptfont0\elevenrm  \scriptscriptfont0\ninerm%
  \textfont1\seventeeni   \scriptfont1\eleveni   \scriptscriptfont1\ninei%
  \textfont2\seventeensy  \scriptfont2\elevensy  \scriptscriptfont2\ninesy%
  \textfont\itfam=\seventeenit\def\it{\fam\itfam\seventeenit}%
  \textfont\bffam=\seventeenbf%
    \scriptfont\bffam=\elevenbf%
      \scriptscriptfont\bffam=\ninebf%
  \def\bf{\fam\bffam\seventeenbf}%
  \textfont\slfam=\seventeensl\def\sl{\fam\slfam\seventeensl}%
  \textfont\ttfam=\seventeentt\def\tt{\fam\ttfam\seventeentt}%
  \textfont\scfam=\seventeencsc\def\sc{\fam\scfam\seventeencsc}%
  \textfont\sffam=\seventeensf\def\sf{\fam\sffam\seventeensf}%
  \textfont\mibfam=\seventeenmib%
  \textfont\sybfam=\seventeensyb%
  \bls{20pt}%
}

\lineskip=1pt      \normallineskip=\lineskip
\lineskiplimit=0pt \normallineskiplimit=\lineskiplimit




\def\Nulle{0}  
\def\Aue{1}    
\def\Afe{2}    
\def\Ace{3}    
\def\Sue{4}    
\def\Hae{5}    
\def\Hbe{6}    
\def\Hce{7}    
\def\Hde{8}    
\def\Kwe{9}    
\def\Txe{10}   
\def\Lie{11}   
\def\Bbe{12}   


\newdimen\DimenA
\newbox\BoxA

\newcount\LastMac \LastMac=\Nulle
\newcount\HeaderNumber \HeaderNumber=0
\newcount\DefaultHeader \DefaultHeader=\HeaderNumber
\newskip\Indent

\newskip\half      \half=5.5pt plus 1.5pt minus 2.25pt
\newskip\one       \one=11pt plus 3pt minus 5.5pt
\newskip\onehalf   \onehalf=16.5pt plus 5.5pt minus 8.25pt
\newskip\two       \two=22pt plus 5.5pt minus 11pt

\def\Half{\vskip-\lastskip\vskip\half}
\def\One{\vskip-\lastskip\vskip\one}
\def\OneHalf{\vskip-\lastskip\vskip\onehalf}
\def\Two{\vskip-\lastskip\vskip\two}


\def\rTenPT{10pt plus \Feathering}

\def\TenPT{10pt plus \Feathering} 
\def\ElevenPT{11pt plus \Feathering}

\def\Raggedright{
 \rightskip=0pt plus \hsize
}

\def\Fullout{
\rightskip=0pt
}

\def\Hang#1#2{
 \hangindent=#1
 \hangafter=#2
}

\def\EveryMac{
 \Fullout
 \everypar{}
}



\def\title#1{
 \EveryMac
 \LastMac=\Nulle
 \global\HeaderNumber=0
 \global\DefaultHeader=1
 \vbox to 1pc{\vss}
 \seventeenpoint
 \Raggedright
 \noindent \bf #1
}

\def\author#1{
 \EveryMac
 \ifnum\LastMac=\Afe \OneHalf
  \else \Two
 \fi
 \LastMac=\Aue
 \fourteenpoint
 \Raggedright
 \noindent \rm #1\par
 \vskip 3pt\relax
}

\def\affiliation#1{
 \EveryMac
 \LastMac=\Afe
 \eightpoint\bls{\TenPT}
 \Raggedright
 \noindent \it #1\par
}

\def\acceptedline#1{
 \EveryMac
 \Two
 \LastMac=\Ace
 \eightpoint\bls{\TenPT}
 \Raggedright
 \noindent \rm #1
}

\def\abstract{%
 \EveryMac
 \Two
 \LastMac=\Sue
 \everypar{\Hang{11pc}{0}}
 \noindent\ninebf ABSTRACT\par
 \tenpoint\bls{\ElevenPT}
 \Fullout
 \noindent\rm
}

\def\keywords{
 \EveryMac
 \Half
 \LastMac=\Kwe
 \everypar{\Hang{11pc}{0}}
 \tenpoint\bls{\ElevenPT}
 \Fullout
 \noindent\hbox{\bf Key words:\ }
 \rm
}


\def\maketitle{%
  \Two%
  \EndOpening%
  \MakePage%
}


\def\pageoffset#1#2{\hoffset=#1\relax\voffset=#2\relax}


\def\Autonumber{
 \global\AutoNumbertrue  
}

\newif\ifAutoNumber \AutoNumberfalse
\newcount\Sec        
\newcount\SecSec
\newcount\SecSecSec

\Sec=0

\def\:{\let\@sptoken= } \:  
\def\:{\@xifnch} \expandafter\def\: {\futurelet\@tempc\@ifnch}

\def\@ifnextchar#1#2#3{%
  \let\@tempMACe #1%
  \def\@tempMACa{#2}%
  \def\@tempMACb{#3}%
  \futurelet \@tempMACc\@ifnch%
}

\def\@ifnch{%
\ifx \@tempMACc \@sptoken%
  \let\@tempMACd\@xifnch%
\else%
  \ifx \@tempMACc \@tempMACe%
    \let\@tempMACd\@tempMACa%
  \else%
    \let\@tempMACd\@tempMACb%
  \fi%
\fi%
\@tempMACd%
}

\def\@ifstar#1#2{\@ifnextchar *{\def\@tempMACa*{#1}\@tempMACa}{#2}}

\def\section{\@ifstar{\@ssection}{\@section}}

\def\@section#1{
 \EveryMac
 \One
 \LastMac=\Hae
 \ninepoint\bls{\ElevenPT}
 \bf
 \Raggedright
 \ifAutoNumber
  \advance\Sec by 1
  \noindent\number\Sec\hskip 1pc \uppercase{#1}
  \SecSec=0
 \else
  \noindent \uppercase{#1}
 \fi
 \nobreak
}

\def\@ssection#1{
 \EveryMac
 \ifnum\LastMac=\Hae \Half
  \else \OneHalf
 \fi
 \LastMac=\Hae
 \tenpoint\bls{\ElevenPT}
 \bf
 \Raggedright
 \noindent\uppercase{#1}
}

\def\subsection#1{
 \EveryMac
 \ifnum\LastMac=\Hae \Half
  \else \OneHalf
 \fi
 \LastMac=\Hbe
 \tenpoint\bls{\ElevenPT}
 \bf
 \Raggedright
 \ifAutoNumber
  \advance\SecSec by 1
  \noindent\number\Sec.\number\SecSec
  \hskip 1pc #1
  \SecSecSec=0
 \else
  \noindent #1
 \fi
 \nobreak
}

\def\subsubsection#1{
 \EveryMac
 \ifnum\LastMac=\Hbe \Half
  \else \OneHalf
 \fi
 \LastMac=\Hce
 \ninepoint\bls{\ElevenPT}
 \it
 \Raggedright
 \ifAutoNumber
  \advance\SecSecSec by 1
  \noindent\number\Sec.\number\SecSec.\number\SecSecSec
  \hskip 1pc #1
 \else
  \noindent #1
 \fi
 \nobreak
}

\def\paragraph#1{
 \EveryMac
 \One
 \LastMac=\Hde
 \ninepoint\bls{\ElevenPT}
 \noindent \it #1
 \rm
}


\def\tx{
 \EveryMac
 \ifnum\LastMac=\Lie \Half\fi
 \ifnum\LastMac=\Hae \nobreak\Half\fi
 \ifnum\LastMac=\Hbe \nobreak\Half\fi
 \ifnum\LastMac=\Hce \nobreak\Half\fi
 \ifnum\LastMac=\Lie \else \noindent\fi
 \LastMac=\Txe
 \ninepoint\bls{\ElevenPT}
 \rm
}


\def\item{
 \par
 \EveryMac
 \ifnum\LastMac=\Lie
  \else \Half
 \fi
 \LastMac=\Lie
 \ninepoint\bls{\ElevenPT}
 \rm
}


\def\bibitem{
 \par
 \EveryMac
 \ifnum\LastMac=\Bbe
  \else \Half
 \fi
 \LastMac=\Bbe
 \Hang{1.5em}{1}
 \eightpoint\bls{\TenPT}
 \Raggedright
 \noindent \rm
}


\newtoks\CatchLine

\def\@journal{Mon.\ Not.\ R.\ Astron.\ Soc.\ }  
\def\@pubyear{1993}        
\def\@pagerange{000--000}  
\def\@volume{000}          
\def\@microfiche{}         %

\def\pubyear#1{\gdef\@pubyear{#1}\@makecatchline}
\def\pagerange#1{\gdef\@pagerange{#1}\@makecatchline}
\def\volume#1{\gdef\@volume{#1}\@makecatchline}
\def\microfiche#1{\gdef\@microfiche{and Microfiche\ #1}\@makecatchline}

\def\@makecatchline{%
  \global\CatchLine{%
    {\rm \@journal {\bf \@volume},\ \@pagerange\ (\@pubyear)\ \@microfiche}}%
}

\@makecatchline 

\newtoks\LeftHeader
\def\shortauthor#1{
 \global\LeftHeader{#1}
}

\newtoks\RightHeader
\def\shorttitle#1{
 \global\RightHeader{#1}
}

\def\PageHead{
 \EveryMac
 \ifnum\HeaderNumber=1 \Pagehead
  \else \Catchline
 \fi
}

\def\Catchline{%
 \vbox to 0pt{\vskip-22.5pt
  \hbox to \PageWidth{\vbox to8.5pt{}\noindent
  \eightpoint\the\CatchLine\hfill}\vss}
 \nointerlineskip
}

\def\Pagehead{%
 \ifodd\pageno
   \vbox to 0pt{\vskip-22.5pt
   \hbox to \PageWidth{\vbox to8.5pt{}\elevenpoint\it\noindent
    \hfill\the\RightHeader\hskip1.5em\rm\folio}\vss}
 \else
   \vbox to 0pt{\vskip-22.5pt
   \hbox to \PageWidth{\vbox to8.5pt{}\elevenpoint\rm\noindent
   \folio\hskip1.5em\it\the\LeftHeader\hfill}\vss}
 \fi
 \nointerlineskip
}

\def\PageFoot{} 

\def\authorcomment#1{%
  \gdef\PageFoot{%
    \nointerlineskip%
    \vbox to 22pt{\vfil%
      \hbox to \PageWidth{\elevenpoint\rm\noindent \hfil #1 \hfil}}%
  }%
}

\everydisplay{\displaysetup}

\newif\ifeqno
\newif\ifleqno

\def\displaysetup#1$${%
 \displaytest#1\eqno\eqno\displaytest
}

\def\displaytest#1\eqno#2\eqno#3\displaytest{%
 \if!#3!\ldisplaytest#1\leqno\leqno\ldisplaytest
 \else\eqnotrue\leqnofalse\def\eqn{#2}\def\eq{#1}\fi
 \generaldisplay$$}

\def\ldisplaytest#1\leqno#2\leqno#3\ldisplaytest{%
 \def\eq{#1}%
 \if!#3!\eqnofalse\else\eqnotrue\leqnotrue
  \def\eqn{#2}\fi}

\def\generaldisplay{%
\ifeqno \ifleqno
   \hbox to \hsize{\noindent
     $\displaystyle\eq$\hfil$\displaystyle\eqn$}
  \else
    \hbox to \hsize{\noindent
     $\displaystyle\eq$\hfil$\displaystyle\eqn$}
  \fi
 \else
 \hbox to \hsize{\vbox{\noindent
  $\displaystyle\eq$\hfil}}
 \fi
}

\def\@notice{%
  \par\Two%
  \bls{12pt}%
  \noindent\tenrm This paper has been produced using the Blackwell
                  Scientific Publications \TeX\ macros.%
}

\outer\def\bye{\@notice\par\vfill\supereject\end}

\everyjob{%
  \Warn{Monthly notices of the RAS journal style (\@typeface)\space
        v\@version,\space \@verdate.}\Warn{}%
}




\newif\if@debug \@debugfalse  

\def\Print#1{\if@debug\immediate\write16{#1}\else \fi}
\def\Warn#1{\immediate\write16{#1}}
\def\wlog#1{}

\newcount\Iteration 

\newif\ifFigureBoxes  
\FigureBoxestrue

\def\Single{0} \def\Double{1}                 
\def\Figure{0} \def\Table{1}                  

\def\InStack{0}  
\def\InZoneA{1}
\def\InZoneB{2}
\def\InZoneC{3}

\newcount\TEMPCOUNT 
\newdimen\TEMPDIMEN 
\newbox\TEMPBOX     
\newbox\VOIDBOX     

\newcount\LengthOfStack 
\newcount\MaxItems      
\newcount\StackPointer
\newcount\Point         
\newcount\NextFigure    
\newcount\NextTable     
\newcount\NextItem      

\newcount\StatusStack   
\newcount\NumStack      
\newcount\TypeStack     
\newcount\SpanStack     
\newcount\BoxStack      

\newcount\ItemSTATUS    
\newcount\ItemNUMBER    
\newcount\ItemTYPE      
\newcount\ItemSPAN      
\newbox\ItemBOX         
\newdimen\ItemSIZE      

\newdimen\PageHeight    
\newdimen\TextLeading   
\newdimen\Feathering    
\newcount\LinesPerPage  
\newdimen\ColumnWidth   
\newdimen\ColumnGap     
\newdimen\PageWidth     
\newdimen\BodgeHeight   
\newcount\Leading       

\newdimen\ZoneBSize  
\newdimen\TextSize   
\newbox\ZoneABOX     
\newbox\ZoneBBOX     
\newbox\ZoneCBOX     

\newif\ifFirstSingleItem
\newif\ifFirstZoneA
\newif\ifMakePageInComplete
\newif\ifMoreFigures \MoreFiguresfalse 
\newif\ifMoreTables  \MoreTablesfalse  

\newif\ifFigInZoneB 
\newif\ifFigInZoneC 
\newif\ifTabInZoneB 
\newif\ifTabInZoneC

\newif\ifZoneAFullPage

\newbox\MidBOX    
\newbox\LeftBOX
\newbox\RightBOX
\newbox\PageBOX   

\newif\ifLeftCOL  
\LeftCOLtrue

\newdimen\ZoneBAdjust

\newcount\ItemFits
\def\Yes{1}
\def\No{2}




\MaxItems=15
\NextFigure=0        
\NextTable=1

\BodgeHeight=6pt
\TextLeading=11pt    
\Leading=11
\Feathering=0pt      
\LinesPerPage=61     
\topskip=\TextLeading
\ColumnWidth=20pc    
\ColumnGap=2pc       

\def\ItemSep{\vskip \TextLeading plus \TextLeading minus 4pt}

\FigureBoxesfalse 

\parskip=0pt
\parindent=18pt
\widowpenalty=0
\clubpenalty=10000
\tolerance=1500
\hbadness=1500
\abovedisplayskip=6pt plus 2pt minus 2pt
\belowdisplayskip=6pt plus 2pt minus 2pt
\abovedisplayshortskip=6pt plus 2pt minus 2pt
\belowdisplayshortskip=6pt plus 2pt minus 2pt

\PageHeight=\TextLeading 
\multiply\PageHeight by \LinesPerPage
\advance\PageHeight by \topskip

\PageWidth=2\ColumnWidth
\advance\PageWidth by \ColumnGap




\newcount\DUMMY \StatusStack=\allocationnumber
\newcount\DUMMY \newcount\DUMMY \newcount\DUMMY
\newcount\DUMMY \newcount\DUMMY \newcount\DUMMY
\newcount\DUMMY \newcount\DUMMY \newcount\DUMMY
\newcount\DUMMY \newcount\DUMMY \newcount\DUMMY
\newcount\DUMMY \newcount\DUMMY \newcount\DUMMY

\newcount\DUMMY \NumStack=\allocationnumber
\newcount\DUMMY \newcount\DUMMY \newcount\DUMMY
\newcount\DUMMY \newcount\DUMMY \newcount\DUMMY
\newcount\DUMMY \newcount\DUMMY \newcount\DUMMY
\newcount\DUMMY \newcount\DUMMY \newcount\DUMMY
\newcount\DUMMY \newcount\DUMMY \newcount\DUMMY

\newcount\DUMMY \TypeStack=\allocationnumber
\newcount\DUMMY \newcount\DUMMY \newcount\DUMMY
\newcount\DUMMY \newcount\DUMMY \newcount\DUMMY
\newcount\DUMMY \newcount\DUMMY \newcount\DUMMY
\newcount\DUMMY \newcount\DUMMY \newcount\DUMMY
\newcount\DUMMY \newcount\DUMMY \newcount\DUMMY

\newcount\DUMMY \SpanStack=\allocationnumber
\newcount\DUMMY \newcount\DUMMY \newcount\DUMMY
\newcount\DUMMY \newcount\DUMMY \newcount\DUMMY
\newcount\DUMMY \newcount\DUMMY \newcount\DUMMY
\newcount\DUMMY \newcount\DUMMY \newcount\DUMMY
\newcount\DUMMY \newcount\DUMMY \newcount\DUMMY

\newbox\DUMMY   \BoxStack=\allocationnumber
\newbox\DUMMY   \newbox\DUMMY \newbox\DUMMY
\newbox\DUMMY   \newbox\DUMMY \newbox\DUMMY
\newbox\DUMMY   \newbox\DUMMY \newbox\DUMMY
\newbox\DUMMY   \newbox\DUMMY \newbox\DUMMY
\newbox\DUMMY   \newbox\DUMMY \newbox\DUMMY

\def\wlog{\immediate\write-1}


\def\GetItemAll#1{%
 \GetItemSTATUS{#1}
 \GetItemNUMBER{#1}
 \GetItemTYPE{#1}
 \GetItemSPAN{#1}
 \GetItemBOX{#1}
}

\def\GetItemSTATUS#1{%
 \Point=\StatusStack
 \advance\Point by #1
 \global\ItemSTATUS=\count\Point
}

\def\GetItemNUMBER#1{%
 \Point=\NumStack
 \advance\Point by #1
 \global\ItemNUMBER=\count\Point
}

\def\GetItemTYPE#1{%
 \Point=\TypeStack
 \advance\Point by #1
 \global\ItemTYPE=\count\Point
}

\def\GetItemSPAN#1{%
 \Point\SpanStack
 \advance\Point by #1
 \global\ItemSPAN=\count\Point
}

\def\GetItemBOX#1{%
 \Point=\BoxStack
 \advance\Point by #1
 \global\setbox\ItemBOX=\vbox{\copy\Point}
 \global\ItemSIZE=\ht\ItemBOX
 \global\advance\ItemSIZE by \dp\ItemBOX
 \TEMPCOUNT=\ItemSIZE
 \divide\TEMPCOUNT by \Leading
 \divide\TEMPCOUNT by 65536
 \advance\TEMPCOUNT by 1
 \ItemSIZE=\TEMPCOUNT pt
 \global\multiply\ItemSIZE by \Leading
}


\def\JoinStack{%
 \ifnum\LengthOfStack=\MaxItems 
  \Warn{WARNING: Stack is full...some items will be lost!}
 \else
  \Point=\StatusStack
  \advance\Point by \LengthOfStack
  \global\count\Point=\ItemSTATUS
  \Point=\NumStack
  \advance\Point by \LengthOfStack
  \global\count\Point=\ItemNUMBER
  \Point=\TypeStack
  \advance\Point by \LengthOfStack
  \global\count\Point=\ItemTYPE
  \Point\SpanStack
  \advance\Point by \LengthOfStack
  \global\count\Point=\ItemSPAN
  \Point=\BoxStack
  \advance\Point by \LengthOfStack
  \global\setbox\Point=\vbox{\copy\ItemBOX}
  \global\advance\LengthOfStack by 1
  \ifnum\ItemTYPE=\Figure 
   \global\MoreFigurestrue
  \else
   \global\MoreTablestrue
  \fi
 \fi
}


\def\LeaveStack#1{%
 {\Iteration=#1
 \loop
 \ifnum\Iteration<\LengthOfStack
  \advance\Iteration by 1
  \GetItemSTATUS{\Iteration}
   \advance\Point by -1
   \global\count\Point=\ItemSTATUS
  \GetItemNUMBER{\Iteration}
   \advance\Point by -1
   \global\count\Point=\ItemNUMBER
  \GetItemTYPE{\Iteration}
   \advance\Point by -1
   \global\count\Point=\ItemTYPE
  \GetItemSPAN{\Iteration}
   \advance\Point by -1
   \global\count\Point=\ItemSPAN
  \GetItemBOX{\Iteration}
   \advance\Point by -1
   \global\setbox\Point=\vbox{\copy\ItemBOX}
 \repeat}
 \global\advance\LengthOfStack by -1
}


\newif\ifStackNotClean

\def\CleanStack{%
 \StackNotCleantrue
 {\Iteration=0
  \loop
   \ifStackNotClean
    \GetItemSTATUS{\Iteration}
    \ifnum\ItemSTATUS=\InStack
     \advance\Iteration by 1
     \else
      \LeaveStack{\Iteration}
    \fi
   \ifnum\LengthOfStack<\Iteration
    \StackNotCleanfalse
   \fi
 \repeat}
}


\def\FindItem#1#2{%
 \global\StackPointer=-1 
 {\Iteration=0
  \loop
  \ifnum\Iteration<\LengthOfStack
   \GetItemSTATUS{\Iteration}
   \ifnum\ItemSTATUS=\InStack
    \GetItemTYPE{\Iteration}
    \ifnum\ItemTYPE=#1
     \GetItemNUMBER{\Iteration}
     \ifnum\ItemNUMBER=#2
      \global\StackPointer=\Iteration
      \Iteration=\LengthOfStack 
     \fi
    \fi
   \fi
  \advance\Iteration by 1
 \repeat}
}


\def\FindNext{%
 \global\StackPointer=-1 
 {\Iteration=0
  \loop
  \ifnum\Iteration<\LengthOfStack
   \GetItemSTATUS{\Iteration}
   \ifnum\ItemSTATUS=\InStack
    \GetItemTYPE{\Iteration}
   \ifnum\ItemTYPE=\Figure
    \ifMoreFigures
      \global\NextItem=\Figure
      \global\StackPointer=\Iteration
      \Iteration=\LengthOfStack 
    \fi
   \fi
   \ifnum\ItemTYPE=\Table
    \ifMoreTables
      \global\NextItem=\Table
      \global\StackPointer=\Iteration
      \Iteration=\LengthOfStack 
    \fi
   \fi
  \fi
  \advance\Iteration by 1
 \repeat}
}


\def\ChangeStatus#1#2{%
 \Point=\StatusStack
 \advance\Point by #1
 \global\count\Point=#2
}



\def\Zone{\InZoneA}

\ZoneBAdjust=0pt

\def\MakePage{
 \global\ZoneBSize=\PageHeight
 \global\TextSize=\ZoneBSize
 \global\ZoneAFullPagefalse
 \global\topskip=\TextLeading
 \MakePageInCompletetrue
 \MoreFigurestrue
 \MoreTablestrue
 \FigInZoneBfalse
 \FigInZoneCfalse
 \TabInZoneBfalse
 \TabInZoneCfalse
 \global\FirstSingleItemtrue
 \global\FirstZoneAtrue
 \global\setbox\ZoneABOX=\box\VOIDBOX
 \global\setbox\ZoneBBOX=\box\VOIDBOX
 \global\setbox\ZoneCBOX=\box\VOIDBOX
 \loop
  \ifMakePageInComplete
 \FindNext
 \ifnum\StackPointer=-1
  \NextItem=-1
  \MoreFiguresfalse
  \MoreTablesfalse
 \fi
 \ifnum\NextItem=\Figure
   \FindItem{\Figure}{\NextFigure}
   \ifnum\StackPointer=-1 \global\MoreFiguresfalse
   \else
    \GetItemSPAN{\StackPointer}
    \ifnum\ItemSPAN=\Single \def\Zone{\InZoneB}\relax
     \ifFigInZoneC \global\MoreFiguresfalse\fi
    \else
     \def\Zone{\InZoneA}
     \ifFigInZoneB \def\Zone{\InZoneC}\fi
    \fi
   \fi
   \ifMoreFigures\Print{}\FigureItems\fi
 \fi
\ifnum\NextItem=\Table
   \FindItem{\Table}{\NextTable}
   \ifnum\StackPointer=-1 \global\MoreTablesfalse
   \else
    \GetItemSPAN{\StackPointer}
    \ifnum\ItemSPAN=\Single\relax
     \ifTabInZoneC \global\MoreTablesfalse\fi
    \else
     \def\Zone{\InZoneA}
     \ifTabInZoneB \def\Zone{\InZoneC}\fi
    \fi
   \fi
   \ifMoreTables\Print{}\TableItems\fi
 \fi
   \MakePageInCompletefalse 
   \ifMoreFigures\MakePageInCompletetrue\fi
   \ifMoreTables\MakePageInCompletetrue\fi
 \repeat
 \ifZoneAFullPage
  \global\TextSize=0pt
  \global\ZoneBSize=0pt
  \global\vsize=0pt\relax
  \global\topskip=0pt\relax
  \vbox to 0pt{\vss}
  \eject
 \else
 \global\advance\ZoneBSize by -\ZoneBAdjust
 \global\vsize=\ZoneBSize
 \global\hsize=\ColumnWidth
 \global\ZoneBAdjust=0pt
 \ifdim\TextSize<23pt
 \Warn{}
 \Warn{* Making column fall short: TextSize=\the\TextSize *}
 \vskip-\lastskip\eject\fi
 \fi
}

\def\MakeRightCol{
 \global\TextSize=\ZoneBSize
 \MakePageInCompletetrue
 \MoreFigurestrue
 \MoreTablestrue
 \global\FirstSingleItemtrue
 \global\setbox\ZoneBBOX=\box\VOIDBOX
 \def\Zone{\InZoneB}
 \loop
  \ifMakePageInComplete
 \FindNext
 \ifnum\StackPointer=-1
  \NextItem=-1
  \MoreFiguresfalse
  \MoreTablesfalse
 \fi
 \ifnum\NextItem=\Figure
   \FindItem{\Figure}{\NextFigure}
   \ifnum\StackPointer=-1 \MoreFiguresfalse
   \else
    \GetItemSPAN{\StackPointer}
    \ifnum\ItemSPAN=\Double\relax
     \MoreFiguresfalse\fi
   \fi
   \ifMoreFigures\Print{}\FigureItems\fi
 \fi
 \ifnum\NextItem=\Table
   \FindItem{\Table}{\NextTable}
   \ifnum\StackPointer=-1 \MoreTablesfalse
   \else
    \GetItemSPAN{\StackPointer}
    \ifnum\ItemSPAN=\Double\relax
     \MoreTablesfalse\fi
   \fi
   \ifMoreTables\Print{}\TableItems\fi
 \fi
   \MakePageInCompletefalse 
   \ifMoreFigures\MakePageInCompletetrue\fi
   \ifMoreTables\MakePageInCompletetrue\fi
 \repeat
 \ifZoneAFullPage
  \global\TextSize=0pt
  \global\ZoneBSize=0pt
  \global\vsize=0pt\relax
  \global\topskip=0pt\relax
  \vbox to 0pt{\vss}
  \eject
 \else
 \global\vsize=\ZoneBSize
 \global\hsize=\ColumnWidth
 \ifdim\TextSize<23pt
 \Warn{}
 \Warn{* Making column fall short: TextSize=\the\TextSize *}
 \vskip-\lastskip\eject\fi
\fi
}

\def\FigureItems{
 \Print{Considering...}
 \ShowItem{\StackPointer}
 \GetItemBOX{\StackPointer} 
 \GetItemSPAN{\StackPointer}
  \CheckFitInZone 
  \ifnum\ItemFits=\Yes
   \ifnum\ItemSPAN=\Single
     \ChangeStatus{\StackPointer}{\InZoneB} 
     \global\FigInZoneBtrue
     \ifFirstSingleItem
      \hbox{}\vskip-\BodgeHeight
     \global\advance\ItemSIZE by \TextLeading
     \fi
     \unvbox\ItemBOX\ItemSep
     \global\FirstSingleItemfalse
     \global\advance\TextSize by -\ItemSIZE
     \global\advance\TextSize by -\TextLeading
   \else
    \ifFirstZoneA
     \global\advance\ItemSIZE by \TextLeading
     \global\FirstZoneAfalse\fi
    \global\advance\TextSize by -\ItemSIZE
    \global\advance\TextSize by -\TextLeading
    \global\advance\ZoneBSize by -\ItemSIZE
    \global\advance\ZoneBSize by -\TextLeading
    \ifFigInZoneB\relax
     \else
     \ifdim\TextSize<3\TextLeading
     \global\ZoneAFullPagetrue
     \fi
    \fi
    \ChangeStatus{\StackPointer}{\Zone}
    \ifnum\Zone=\InZoneC \global\FigInZoneCtrue\fi
  \fi
   \Print{TextSize=\the\TextSize}
   \Print{ZoneBSize=\the\ZoneBSize}
  \global\advance\NextFigure by 1
   \Print{This figure has been placed.}
  \else
   \Print{No space available for this figure...holding over.}
   \Print{}
   \global\MoreFiguresfalse
  \fi
}

\def\TableItems{
 \Print{Considering...}
 \ShowItem{\StackPointer}
 \GetItemBOX{\StackPointer} 
 \GetItemSPAN{\StackPointer}
  \CheckFitInZone 
  \ifnum\ItemFits=\Yes
   \ifnum\ItemSPAN=\Single
    \ChangeStatus{\StackPointer}{\InZoneB}
     \global\TabInZoneBtrue
     \ifFirstSingleItem
      \hbox{}\vskip-\BodgeHeight
     \global\advance\ItemSIZE by \TextLeading
     \fi
     \unvbox\ItemBOX\ItemSep
     \global\FirstSingleItemfalse
     \global\advance\TextSize by -\ItemSIZE
     \global\advance\TextSize by -\TextLeading
   \else
    \ifFirstZoneA
    \global\advance\ItemSIZE by \TextLeading
    \global\FirstZoneAfalse\fi
    \global\advance\TextSize by -\ItemSIZE
    \global\advance\TextSize by -\TextLeading
    \global\advance\ZoneBSize by -\ItemSIZE
    \global\advance\ZoneBSize by -\TextLeading
    \ifFigInZoneB\relax
     \else
     \ifdim\TextSize<3\TextLeading
     \global\ZoneAFullPagetrue
     \fi
    \fi
    \ChangeStatus{\StackPointer}{\Zone}
    \ifnum\Zone=\InZoneC \global\TabInZoneCtrue\fi
   \fi
  \global\advance\NextTable by 1
   \Print{This table has been placed.}
  \else
  \Print{No space available for this table...holding over.}
   \Print{}
   \global\MoreTablesfalse
  \fi
}


\def\CheckFitInZone{%
{\advance\TextSize by -\ItemSIZE
 \advance\TextSize by -\TextLeading
 \ifFirstSingleItem
  \advance\TextSize by \TextLeading
 \fi
 \ifnum\Zone=\InZoneA\relax
  \else \advance\TextSize by -\ZoneBAdjust
 \fi
 \ifdim\TextSize<3\TextLeading \global\ItemFits=\No
 \else \global\ItemFits=\Yes\fi}
}

\def\BF#1#2{
 \ItemSTATUS=\InStack
 \ItemNUMBER=#1
 \ItemTYPE=\Figure
 \if#2S \ItemSPAN=\Single
  \else \ItemSPAN=\Double
 \fi
 \setbox\ItemBOX=\vbox{}
}

\def\BT#1#2{
 \ItemSTATUS=\InStack
 \ItemNUMBER=#1
 \ItemTYPE=\Table
 \if#2S \ItemSPAN=\Single
  \else \ItemSPAN=\Double
 \fi
 \setbox\ItemBOX=\vbox{}
}

\def\BeginOpening{%
 \hsize=\PageWidth
 \global\setbox\ItemBOX=\vbox\bgroup
}

\let\begintopmatter=\BeginOpening  

\def\EndOpening{%
 \egroup
 \ItemNUMBER=0
 \ItemTYPE=\Figure
 \ItemSPAN=\Double
 \ItemSTATUS=\InStack
 \JoinStack
}


\newbox\tmpbox

\def\FC#1#2#3#4{%
  \ItemSTATUS=\InStack
  \ItemNUMBER=#1
  \ItemTYPE=\Figure
  \if#2S
    \ItemSPAN=\Single \TEMPDIMEN=\ColumnWidth
  \else
    \ItemSPAN=\Double \TEMPDIMEN=\PageWidth
  \fi
  {\hsize=\TEMPDIMEN
   \global\setbox\ItemBOX=\vbox{%
     \ifFigureBoxes
       \B{\TEMPDIMEN}{#3}
     \else
       \vbox to #3{\vfil}%
     \fi%
     \eightpoint\rm\bls{\rTenPT}%
     \vskip 5.5pt plus 6pt%
     \setbox\tmpbox=\vbox{#4\par}%
     \ifdim\ht\tmpbox>10pt 
       \noindent #4\par%
     \else
       \hbox to \hsize{\hfil #4\hfil}%
     \fi%
   }%
  }%
  \JoinStack%
  \Print{Processing source for figure {\the\ItemNUMBER}}%
}


\def\TH#1#2#3#4{%
 \ItemSTATUS=\InStack
 \ItemNUMBER=#1
 \ItemTYPE=\Table
 \if#2S \ItemSPAN=\Single \TEMPDIMEN=\ColumnWidth
  \else \ItemSPAN=\Double \TEMPDIMEN=\PageWidth
 \fi
{\hsize=\TEMPDIMEN
\eightpoint\bls{\rTenPT}\rm
\global\setbox\ItemBOX=\vbox{\noindent#3\vskip 5.5pt plus5.5pt\noindent#4}}
 \JoinStack
 \Print{Processing source for table {\the\ItemNUMBER}}
}


\def\UnloadZoneA{%
\FirstZoneAtrue
 \Iteration=0
  \loop
   \ifnum\Iteration<\LengthOfStack
    \GetItemSTATUS{\Iteration}
    \ifnum\ItemSTATUS=\InZoneA
     \GetItemBOX{\Iteration}
     \ifFirstZoneA \vbox to \BodgeHeight{\vfil}%
     \FirstZoneAfalse\fi
     \unvbox\ItemBOX\ItemSep
     \LeaveStack{\Iteration}
     \else
     \advance\Iteration by 1
   \fi
 \repeat
}

\def\UnloadZoneC{%
\Iteration=0
  \loop
   \ifnum\Iteration<\LengthOfStack
    \GetItemSTATUS{\Iteration}
    \ifnum\ItemSTATUS=\InZoneC
     \GetItemBOX{\Iteration}
     \ItemSep\unvbox\ItemBOX
     \LeaveStack{\Iteration}
     \else
     \advance\Iteration by 1
   \fi
 \repeat
}


\def\ShowItem#1{
  {\GetItemAll{#1}
  \Print{\the#1:
  {TYPE=\ifnum\ItemTYPE=\Figure Figure\else Table\fi}
  {NUMBER=\the\ItemNUMBER}
  {SPAN=\ifnum\ItemSPAN=\Single Single\else Double\fi}
  {SIZE=\the\ItemSIZE}}}
}

\def\ShowStack{%
 \Print{}
 \Print{LengthOfStack = \the\LengthOfStack}
 \ifnum\LengthOfStack=0 \Print{Stack is empty}\fi
 \Iteration=0
 \loop
 \ifnum\Iteration<\LengthOfStack
  \ShowItem{\Iteration}
  \advance\Iteration by 1
 \repeat
}

\def\B#1#2{%
\hbox{\vrule\kern-0.4pt\vbox to #2{%
\hrule width #1\vfill\hrule}\kern-0.4pt\vrule}
}

\def\Ref#1{\begingroup\global\setbox\TEMPBOX=\vbox{\hsize=2in\noindent#1}\endgroup
\ht1=0pt\dp1=0pt\wd1=0pt\vadjust{\vtop to 0pt{\advance
\hsize0.5pc\kern-10pt\moveright\hsize\box\TEMPBOX\vss}}}

\def\MarkRef#1{\leavevmode\thinspace\hbox{\vrule\vtop
{\vbox{\hrule\kern1pt\hbox{\vphantom{\rm/}\thinspace{\rm#1}%
\thinspace}}\kern1pt\hrule}\vrule}\thinspace}%


\output{%
 \ifLeftCOL
  \global\setbox\LeftBOX=\vbox to \ZoneBSize{\box255\unvbox\ZoneBBOX}
  \global\LeftCOLfalse
  \MakeRightCol
 \else
  \setbox\RightBOX=\vbox to \ZoneBSize{\box255\unvbox\ZoneBBOX}
  \setbox\MidBOX=\hbox{\box\LeftBOX\hskip\ColumnGap\box\RightBOX}
  \setbox\PageBOX=\vbox to \PageHeight{%
  \UnloadZoneA\box\MidBOX\UnloadZoneC}
  \shipout\vbox{\PageHead\box\PageBOX\PageFoot}
  \global\advance\pageno by 1
  \global\HeaderNumber=\DefaultHeader
  \global\LeftCOLtrue
  \CleanStack
  \MakePage
 \fi
}


\catcode `\@=12 


\pageoffset{-2.5pc}{0pc}


\Autonumber  


\pagerange{}
\pubyear{}
\volume{}

\begintopmatter  

\title{Microlensing by the Milky Way Halo}
\author{N.W. Evans$^{1,2}$ and J. Jijina$^3$}
\affiliation{$^1$Department of Mathematics, Massachusetts Institute
of Technology, Cambridge, Massachusetts, MA 02139, USA}
\affiliation{$^2$Theoretical Physics, Department of Physics,
1 Keble Rd, Oxford, OX1 3NP, UK}
\affiliation{$^3$Department of Physics, University of California,
San Diego, California 92093, USA}

\shortauthor{N.W. Evans and J. Jijina}
\shorttitle{Microlensing by the Milky Way Halo}


\acceptedline{}

\abstract The measurements of the possible gravitational microlensing
events are analysed with a simple yet accurate disc--halo model of the
Milky Way Galaxy. This comprises a luminous exponential disc embedded in a
flattened dark matter halo with density varying like ${\rm distance}^{-1.8}$.
Including a disc has the important effect of lowering the implied masses
of the dark matter objects. For the possible detection reported by Alcock
et al (1993), the inferred mass of the lens lies in the range
$\sim 0.01 - 0.15 M_\odot$. The candidate events of Aubourg et
al (1993) have slightly larger implied masses of $\sim 0.025 - 0.35\ M_\odot$
and $\sim 0.03 - 0.45\ M_\odot$ respectively. These are consistent
with the deflecters being either brown dwarfs or low mass stars.
If there is no disc dark matter and the halo is completely composed of
baryonic dark objects of typical mass $ \sim 0.08\ M_\odot$, then the
monitoring of $1.8 \times 10^6$ stars in the Large Magellanic Cloud
will provide at least $8-9$ detections a year, in the limit of $100 \%$
efficiency.

\keywords Galaxy : halo -- dark matter -- gravitational lensing

\maketitle  

\section{Introduction}

\tx Paczy\'nski (1986) made the influential proposal that dark objects
-- possibly brown dwarfs, low mass stars or black holes -- in the halo of
the Milky Way can act as gravitational microlenses, causing the
occasional amplification of the images of stars in the Magellanic
Clouds. He used the simplest possible model of a galactic halo -- the
singular isothermal sphere -- to estimate the order--of--magnitude
effects. Griest (1991) refined the model by introducing a core
radius and analysing the importance of relative motion of observer
and source. Two groups (Alcock et al 1993; Aubourg et al 1993) have now
claimed possible detections of microlensing events of Large
Magellanic Cloud stars. Evidently, this warrants the development of a
more accurate representation of the Milky Way galaxy to enable a
sophisticated confrontation with the observations. Such is the purpose
of this Letter.

\section{A Disc--Halo Model for the Milky Way}

\tx The observables that can be measured by the experimental programs
are the microlensing rate $\Gamma$ and the average duration of events
$\langle t_e \rangle$. To compare the predictions of a theoretical
model with the data is only possible if the velocity distribution of
the massive compact halo objects (henceforth Machos) is known. As the
Machos are collisionless, the distribution function must obey Jeans'
theorem and depend only on the isolating integrals of the motion
(see e.g., Binney \& Tremaine 1987, p. 220).

Let us assume that the candidate microlensing events are caused
by Machos in the halo of the Milky Way. This remains the most plausible
location, although a number of investigators (Gould, Miralda--Escud\'e
\& Bahcall 1993; Giudice, Mollerach, \& Roulet 1994) have recently
suggested that thin or thick discs and spheroids of dark matter
objects are viable alternatives. Given our hypothesis, the most
important thing is to reproduce the structure of the dark matter halo
as accurately as possible at the locations where typical microlensing
events occur -- between $10-40\ {\rm kpc}$ along the line--of--sight
from the observer to the Large Magellanic Cloud ($\ell = 280^{\rm o},
b = -33^{\rm o}$). Our model of the Milky Way has just two components
-- a luminous thin exponential disc and a flattened dark matter halo.
This follows the lead provided by Gilmore, Wyse \& Kuijken (1989), who
claim that there is no evidence for dark matter in the disc. The density
of the bulge and spheroidal component is insignificant at the radii of
interest and can be neglected. But, it is important to include the disc --
its contribution to the rotation curve at the solar radius is $\sim
40 \% $ and this alters the normalisation of the halo!

Let us adopt the simple and reasonable assumption that the galactic
rotation curve is fairly flat out to the Large Magellanic Cloud and
deconvolve this into contributions from the visible disc and dark
matter halo. From stellar kinematic analyses, the local surface density
of observable disc matter is $\sim 48\ M_\odot\ {\rm pc}^{-2}$
(Gilmore, Wyse \& Kuijken 1989). The Milky Way disc -- like those of
external spiral galaxies -- is exponential in radius. The scale--length
is uncertain, but a typical estimate is $3.5\ {\rm kpc}$ (de Vaucouleurs
\& Pence, 1978).

The equipotentials of dark matter haloes are roughly stratified on similar
concentric spheroids. A good approximation to their gravitational potential
is given by (Evans, 1994)
$$\psi = {v_0^2 R_c^\beta/ \beta \over (R_c^2 + R^2 + z^2q^{-2})^{\beta/2}}
,\qquad \beta \neq 0,\eqno(1)$$
where ($R,z$) are cylindrical polar coordinates. Here, $R_c$ is the core
radius of the halo and $q$ is the axis ratio of the equipotentials. So,
$q =1$ corresponds to a spherical halo, while $q <1$ implies the
mass model is flattened. The parameter $v_0$ is a velocity that measures
the central depth of the potential well. The halo mass density is
$$\rho  = {v_0^2 R_c^\beta[ R_c^2 (1 + 2 q^2) +
R^2(1-\beta q^2) + z^2(2-{1 + \beta\over q^2})]
\over 4\pi Gq^2 (R_c^2 + R^2 + z^2q^{-2})^{(\beta +4)/2}},\eqno(2)$$
which falls off like $r^{-2-\beta}$ at large radii. The models --
known as the power--law haloes -- are useful because the self--consistent
phase space distribution function is known analytically and has the
simple form
$$F = AL_z^2|E|^{4/\beta - 3/2} + B|E|^{4/\beta - 1/2} + C|E|^{2/\beta
-1/2},\eqno(3)$$
where $A, B$ and $C$ are constants given in Evans (1994). Note that,
in accord with Jeans' theorem, the distribution function depends on
the two isolating integrals of motion, the binding energy $E$ and
the angular momentum component parallel to the symmetry axis $L_z$.
Strictly speaking, the distribution of velocities that builds the
halo in the combined potential field of both disc and halo is what
we need -- however, the potential of the disc is negligible compared
to that of the halo at the heights above the galactic plane where
typical microlensing events occur and so the self--consistent solution
(2) is an excellent approximation. For example, at a typical
microlens location of half--way along the line--of--sight towards
the Large Magellanic Cloud, the percentage error in neglecting the
contribution of the disc to the total gravitational potential is
$< 4 \%$.

Although the core radius of the halo $R_c$ is not well--known, our
results are insensitive to this parameter. A reasonable value is
$\sim 2\ {\rm kpc}$ (Bahcall, Schmidt \& Soneira 1983). The axis ratio
of the equipotentials $q$ controls the flattening of the model. Evidence
on the ellipticity of the dark halo is sketchy. All we know for sure
is that $N$--body simulations of gravitational collapse (e.g.,
Dubinski \& Carlberg 1991) invariably produce flattened dark haloes. We
shall give results for the extremes of E0 and E6 haloes.
The galactic constants -- the solar radius $R_0$ and the local
circular speed $v_{circ}(R_0)$ -- are uncertain to at least $10 \%$.
Merrifield (1992) analysed the data on the thickness of the HI
layer and concluded $R_0 = 7.9\ {\rm kpc}$ and $v_{circ}(R_0)
= 200\ {\rm kms}^{-1}$. This is in accord with a number of recent
investigations (Rohlfs et al 1986; Reid 1989) which have reported
that the IAU estimates (Kerr \& Lynden--Bell 1986) of
$R_0 = 8.5\ {\rm kpc}$ and $v_{circ}(R_0) = 220\ {\rm kms}^{-1}$ are
too high. Note that increasing the local circular speed implies a
larger halo density and so a greater number of microlensing events.
As part of our aim is to estimate the minimum number of detections
consistent with a halo built from Machos, it makes sense to adopt
Merrifield's data--points. The remaining parameters $v_0$ and
$\beta$ in the power--law halo (1) are chosen to ensure that the
combined rotation curve of disc and halo is flattish and
$\sim 200\ {\rm kms}^{-1}$ outwards from the solar circle.
This yields $v_0 = 138\ {\rm kms}^{-1}$ and $\beta = -0.2$.

So, our Milky Way model has a luminous thin exponential disc embedded
in a flattened dark matter halo with a simple distribution function.
The mass density of the dark halo varies like ${\rm distance}^{-1.8}$ at
large radii (c.f., Bahcall, Schmidt \& Soneira 1983). The model balances
simplicity with realism.

\section{The Microlensing Rates and the Frequency Distributions}

\tx Our aim in this section is to evaluate the observables -- the microlensing
rate $\Gamma$ and average duration $\langle t_e \rangle$ -- together with
the frequency distributions of events. In the light of these calculations,
we analyse the data on the possible detections in Alcock et al (1993) and
Aubourg et al (1993).

The amplification $A$ of the stellar image is related to the impact
parameter $u$ of the Macho in units of the Einstein ring radius $R_E$
by (e.g., Refsdal 1964; Paczy\'nski 1986)
$$A(u) = {u^2 +2\over u (u^2 +4)^{1/2} }.\eqno(4)$$
Let us assume that an observing program can detect events above a threshold
amplification $A_T$ and impact parameter $u_T$. Then, an event occurs
whenever a Macho lies within a microlensing tube with circular cross--section
of radius $u_TR_E$ that connects the eye of the observer to the
source of radiation in the Large Magellanic Cloud. The fatness of the
tube varies because the threshold impact parameter for the deflecting
Macho to cause detectable image amplification changes with distance along
the line--of--sight. The rate is the number of Machos per unit time entering
the tube. In reality, the tube sweeps slowly through the Milky Way as both
the observer and the source are in motion. Griest (1991) examined the
effects of this relative motion and concluded that it leads to only a
slight increment in the total rate for the microlensing of Large Magellanic
Cloud stars. This justifies the approximation of stationary observer and
source that is used below.

Let us suppose the radiation source is at a distance $L$ from the
sun and has galactic coordinates ($\ell,b$). By considering an
element of the microlensing tube of angular extent $u_TR_E d\alpha$
and length $ds$ along the line--of--sight, the total rate is (Griest
1991)
$$\Gamma = {u_T\over M} \int Fv_r^2 \cos \theta R_E dv_r dv_s d\theta
d\alpha ds,\eqno(5)$$
where $M$ is the typical mass of the Machos and $v_r dv_rdv_s d\theta$
is the volume element in cylindrical polar coordinates in velocity space.
For the Large Magellanic Cloud, the rate is $1.35 \times 10^{-6} u_T/
(M/M_\odot)^{1/2}\ {\rm event\ yr}^{-1}$ if the Milky Way halo is spherical,
and $1.41 \times 10^{-6} u_T/(M/M_\odot)^{1/2}\ {\rm event\ yr}^{-1}$ if it
is as flat as E6. The second observable is the average duration of events
$\langle t_e \rangle$. This is the ratio of the number of Machos in the
microlensing tube $\tau$ to the rate at which they enter $\Gamma$, i.e.,
$$\langle t_e \rangle = {\tau\over \Gamma},\eqno(6)$$
Now, the number of Machos in the tube -- or the optical depth -- is
$$\tau = {1\over M} \int_0^L \pi u_T^2R_E^2 \rho ds.\eqno(7)$$
Again for the Large Magellanic Cloud, the average event duration is
$0.348 u_T (M/M_\odot)^{1/2}\ {\rm yr}$ for an E0 halo, and $0.368
u_T (M/M_\odot)^{1/2}\ {\rm yr}$ for an E6 halo. Exact expressions
for the microlensing rate and optical depth in the power--law haloes
are given in Griest et al (in preparation).

Let us emphasise that it is not possible to deduce unambiguously the
typical mass of a deflector from the data presently available. The best
that can be done is to estimate the most probable mass $M$ that caused
an event of duration $t_e$ above the threshold $u_T$. The frequency
distribution of events $p_M(t_e) = d\Gamma / dt_e$ is evaluated by
changing integration variables in (5) from $\theta$ to $t_e$
$$p_M(t_e) = {d\Gamma \over dt_e} = {1\over M}\int {F\over \tan \theta}
 v_r^3 dv_r dv_s d\alpha ds.\eqno(8)$$
The normalised distribution of events as a function of duration is shown
in figure 2. The graph is plotted for a threshold $u_T =1$ and Macho mass
of $0.08 M_\odot$. For other values, the $t_e$ axis is scaled by
$u_TM^{1/2}$ and the rate axis by $1/(u_TM^{1/2})$. As Griest (1991)
argued, the most likely mass giving rise to an event of typical
duration $\langle t_e \rangle$ is the value of $M$ for which $p_M(t_e)$
is largest. This enables us to generate a curve of relative probabilities
of Macho masses that might have caused an event of prescribed duration.

The candidate event detected by Alcock et al (1993) had a duration of
$t_e = 0.092\ {\rm yr}$ and a peak amplification of $A = 6.86$. The high
amplification implies that the Macho passed close to the centre of the
microlensing tube -- the impact parameter is $0.15 R_E$. If $v$ denotes
the Macho velocity, then the duration of the event is $1.98 R_E/ v$.
The typical duration $\langle t_e \rangle$ is an average over all
amplifications greater than the threshold and so is $\sim \pi R_E/ (2 v)$.
Therefore, the average duration $\langle t_e \rangle$ implied by the
candidate event is roughly $0.073\ {\rm yr}$. (This effect was first
noted in Monte Carlo simulations by W. Sutherland (1993, private
communication)). Figure 3 shows the relative probability of Macho masses
for a spherical halo (full line) and an E6 halo (dashed line), from which
we deduce the most probable mass of $\sim 0.04$ or $0.03 \ M_\odot$
depending on whether the halo is E0 or E6. Masses of $0.01\ M_\odot$
and $0.15\ M_\odot$ are roughly half as likely.

Now, the end of the main sequence is generally taken as $0.08\ M_\odot$
(e.g., Nelson, 1990). Objects with masses larger than this are
dim stars, objects smaller are brown dwarfs. The fates of these
two classes of object are quite different -- low mass stars achieve
a state of almost complete thermal equilibrium on the order of a
Hubble time, whereas brown dwarfs are destined to cool to a fully
degenerate configuration. The only fair conclusion from figure 3 is that
the cause of the candidate event reported by Alcock et al (1993) might
be microlensing by a dark halo object with a mass in the range
$0.01 - 0.15\ M_\odot$. This is consistent with both brown dwarfs
and low mass stars.  It is substantially smaller than the
$0.03 - 0.5\ M_\odot$ range reported by Alcock et al (1993).

The EROS collaboration uses a different definition of characteristic
time. To analyse their data, we first work out the timescale over which
the amplification exceeded $A = 1.34$ for their events. This gives
durations of $0.134\ {\rm yr}$ and $0.154\ {\rm yr}$ respectively for
their first and second detection possibilities. Again, we now use the
reported amplifications of $2.51$ and $3.02$ to convert to average
durations $\langle t_e \rangle$ of $0.116\ {\rm yr}$ and $0.129\ {\rm yr}$
respectively. The first event is consistent with a range of $0.025 - 0.35
\ M_\odot$, with the most likely value $\sim 0.10\ M_\odot$. For the
second event, the most likely mass is $\sim 0.12\ M_\odot$, with masses
of $0.03\ M_\odot$ and $0.45\ M_\odot$ half as likely.

Our final aim is to estimate the {\it minimum} number of detections that
the microlensing programs should find -- assuming the halo is completely
composed of Machos. An important source of uncertainty lies in the value
of the local circular speed $v_{circ}(R_0)$, which alters the normalisation
of the halo model. If a higher circular speed is postulated, then the
density of Machos required to maintain this value out to the Large
Magellanic Cloud is correspondingly greater, implying an increased number
of detections. This is why we have chosen the estimate of Merrifield (1992),
which is amongst the lowest of recent determinations. Let us assume that
the Machos have typical masses of the order mass $\sim 0.08\ M_\odot$.
The team of Alcock et al (1991) reported monitoring $1.8 \times 10^6$
stars, and so the annual number of detections in the limit of $100 \%$
efficiency is $\sim 8$ or $9$, depending on whether the halo is round
or E6. This is the smallest conceivable rate consistent with solely
baryonic dark matter. This conclusion depends strongly on our assumption
of the absence of dark matter in the disc, and more weakly on the
errors in the Galactic constants (see Griest at al (in preparation) for
a longer discussion of the microlensing rate scatter).

\section{Conclusions}

\tx Microlensing observables are determined by the structure of the halo
in velocity space. So, we have provided a simple and accurate disc--halo
model of the Milky Way -- with distribution functions for the dark
matter. This is a good representation of the known structure of our Galaxy.
However, we caution that other models -- with rather different
lensing properties, such as thick discs of dark matter -- are also
compatible with our existing knowledge.

Our inferred masses for the lensing objects are less than those reported
by other investigators (see e.g., Alcock et al 1993, Aubourg et al 1993
and Jetzner \& Masso 1993). This has a simple explanation. In our
model, part of the local centrifugal balance is provided by the gravity
field of the disc. The potential deep in the halo is {\it lowered} in
comparison with models like the isothermal sphere, where the local
centrifugal balance is provided by the halo alone. A lower potential means
that the Machos move on average slower, and so smaller masses are needed
to give the observed event durations. Our most likely values for the masses
straddle the boundary between low mass stars and brown dwarfs. The
data is consistent with microlensing by both possibilities. Stronger
conclusions must await more detections.

If the halo is completely composed of objects of mass $\sim 0.08\
M_\odot$, the expected annual number of detections for a progam
monitoring $1.8 \times 10^6$ stars is $\sim 8 -9$ in the limit of
$100 \% $ efficiency. The actual efficiency of the detectors is under
investigation at the moment by the MACHO group. The EROS
collaboration have estimated the efficiency of their equipment
to be approximately $50 \%$ (Aubourg et al 1993). This suggests
that the annual rate of detections should be at least $4$ or so
if the halo is entirely built of Machos.

\section*{Acknowledgments}

\tx NWE and JJ are especially indebted to Kim Griest for generous advice
and discussions, which aided our understanding and enriched the paper.
NWE also thanks James Binney and Subir Sarkar for asking useful
questions. His research is supported by the Science and Engineering
Research Council of the United Kingdom. JJ acknowledges support from
Kim Griest's US Department of Energy OJI award.

\section*{References}

\bibitem Alcock, C. et al, Nature, 1993, 365, 621
\bibitem Aubourg, E. et al, Nature, 1993, 365, 623
\bibitem Bahcall, J., Schmidt, M., \& Soneira, R., 1983, ApJ, 265, 730
\bibitem Binney, J.J., \& Tremaine, S.D., 1987, Galactic Dynamics,
Princeton University Press, Princeton
\bibitem de Vaucouleurs, G., \& Pence, W.D., 1978, AJ, 83, 1163
\bibitem Dubinski, J., \& Carlberg, R., 1991, ApJ, 378, 496
\bibitem Evans, N.W., 1994, MNRAS, in press
\bibitem Gilmore, G., Wyse, R.F.G., \& Kuijken, K., 1989, ARAA,
27, 555
\bibitem Giudice, G., Mollerach, S., \& Roulet, E., 1994, preprint
\bibitem Gould, A., Miralda--Escud\'e, J., \& Bahcall, J., 1993,
preprint
\bibitem Griest, K., 1991, ApJ, 366, 412
\bibitem Jetzer, P., \& Masso, E., 1993, Phys. Lett. B, in press
\bibitem Kerr, F.J., \& Lynden--Bell, D., 1986, MNRAS, 221, 1023
\bibitem Merrifield, M., 1992, AJ, 103, 1552
\bibitem Nelson, L.A., 1990, in Baryonic Dark Matter, eds. Lynden--Bell,
D., \& Gilmore, G., Dordrecht, Reidel, p. 67
\bibitem Paczy\'nski, B., 1986, ApJ, 304, 1
\bibitem Reid, M.J., 1989, in The Center of the Galaxy, IAU
Symposium No. 136, ed. Morris, M., Kluwer, Dordrecht
\bibitem Refsdahl, S., 1964, MNRAS, 128, 295
\bibitem Rohlfs, K., Chini, R., Wink, J.E., \& B\"ohme, R., 1986,
Astr. Ap., 201, 51

\section{Figures}

\noindent
Figure 1: Plot of the contributions of the exponential
disc and the power--law halo to the total rotation curve of our model
of the Milky Way Galaxy

\noindent
Figure 2: The normalised distribution of microlensing event
rate as a function of event duration for a source in the Large Magellanic
Cloud. The deflecting lens is taken to be a brown dwarf or low mass star
of $0.08 M_\odot$. The full line refers to an E0 halo, the broken line
to an E6 halo.

\noindent
Figure 3: The relative probabilities of Macho mass giving rise
to events of average duration $\langle t_e \rangle = 0.073\ {\rm yr}$. The
full line refers to an E0 halo, the broken line to an E6 halo.

\end